\documentclass[12pt]{article}
\usepackage{epsfig}
\usepackage{graphicx}
\usepackage{a4}
\usepackage{latexsym}
\usepackage{cite}

\usepackage{pslatex}
\usepackage[latin1]{inputenc}
\usepackage[T1]{fontenc}

\newcommand{\beq}{\begin{equation}}
\newcommand{\eeq}{\end{equation}}
\newcommand{\bea}{\begin{eqnarray}}
\newcommand{\eea}{\end{eqnarray}}
\newcommand{\nn}{\nonumber}
\newcommand{\ra}{\rightarrow}
\newcommand{\MSb}{$\overline{\mbox{MS}}$}
\newcommand{\as}{\alpha_{\rm s}}
\newcommand{\ars}{a_{\rm s}}
\newcommand{\ep}{\epsilon}
\newcommand{\DD}{{\cal D}}
\newcommand{\ec}{\gamma_e}
\newcommand{\ecs}{\gamma_e^{\,2}}
\newcommand{\ect}{\gamma_e^{\,3}}
\newcommand{\ecf}{\gamma_e^{\,4}}

\def\z#1{{\zeta_{#1}}}
\def\zs{{\zeta_{2}^{\,2}}}

\def\ca{{C^{}_A}}
\def\cas{{C^{\,2}_A}}
\def\cat{{C^{\,3}_A}}
\def\cf{{C^{}_F}}
\def\cfs{{C^{\, 2}_F}}
\def\cft{{C^{\, 3}_F}}
\def\nf{{n^{}_{\! f}}}
\def\nsq{{n^{\,2}_{\! f}}}

\def\gz{\gamma_0^{}}
\def\go{\gamma_1^{}}
\def\gt{\gamma_2^{}}

\def\co{c_1^{}}
\def\ct{c_2^{}}
\def\cd{c_3^{}}
\def\ao{a_1^{}}
\def\at{a_2^{}}
\def\bo{b_1^{}}

\begin{document}
\setlength{\parskip}{0.2cm}
\setlength{\baselineskip}{0.55cm}

\begin{titlepage}
\noindent
DESY 05-152, SFB/CPP-05-46 \hfill {\tt hep-ph/0508265}\\
DCPT/05/110, $\,\,$IPPP/05/55\\[1mm]
August 2005 \\
\vspace{1.8cm}
\begin{center}
\Large
{\bf Higher-Order Soft Corrections to }  \\[2mm]
{\bf Lepton Pair and Higgs Boson Production} \\
\vspace{2.2cm}
\large
S. Moch$^{\, a}$ and A. Vogt$^{\, b}$\\
\vspace{1.4cm}
\normalsize
{\it $^a$Deutsches Elektronensynchrotron DESY \\
\vspace{0.1cm}
Platanenallee 6, D--15735 Zeuthen, Germany}\\
\vspace{0.5cm}
{\it $^b$IPPP, Department of Physics, University of Durham\\
\vspace{0.1cm}
South Road, Durham DH1 3LE, United Kingdom}\\
\vfill
\large
{\bf Abstract}
\vspace{-0.2cm}
\end{center}
Utilizing recent three-loop results on the quark and gluon splitting functions 
and form factors, we derive the complete threshold-enhanced third-order 
(N$^3$LO) QCD corrections to the total cross sections for the production of 
lepton pairs and the Higgs boson in hadron collisions. These results, for the 
latter case obtained in the heavy top-quark limit, are employed to extend the 
threshold resummation for these processes to the fourth logarithmic order. 
We investigate the numerical impact of the higher-order corrections for Higgs 
boson production at the {\sc Tevatron} and the LHC. Our results, suitably
treated in Mellin $N$-space, provide a sufficiently accurate approximation to 
the full N$^3$LO contributions. Corrections of about 5\% at the LHC and 10\% at 
the {\sc Tevatron} are found for typical Higgs masses. The N$^3$LO predictions 
exhibit a considerably reduced dependence on the renormalization scale with,
for the first time, stationary points close to the Higgs mass. 
\vspace{1.0cm}
\end{titlepage}

% -----------------------------------------------------------------------------

\noindent
The production of lepton pairs and especially the Higgs boson $H$, see Refs.\
\cite{Djouadi:2005gi} for detailed reviews, are among the most important 
processes in high-energy proton collisions. The corresponding cross sections 
receive sizable higher-order QCD corrections, necessitating calculations beyond
the standard next-to-leading order (NLO) of perturbative QCD. 
After the early calculation of the next-to-next-to-leading order (NNLO) 
corrections to the Drell-Yan process in Refs.~\cite
{Matsuura:1989sm,Hamberg:1991np}, considerable progress has been achieved in 
this field during the last five years. The NNLO corrections are now completely 
known also for Higgs production in the heavy top-quark limit
\cite{Harlander:2001is,Catani:2001ic,Harlander:2002wh,Anastasiou:2002yz,%
Ravindran:2003um,Ravindran:2004mb,Anastasiou:2005qj}, and the all-order 
resummation of the threshold logarithms~\cite{Sterman:1987aj,Catani:1989ne} has
been extended to the next-to-next-to-leading logarithmic (NNLL) accuracy for 
both processes \cite{Vogt:2000ci,Catani:2003zt}. Still, at colliders energies 
especially for Higgs production, corrections of yet higher order are not 
entirely negligible. 

In this letter we derive the complete logarithmically enhanced soft-emission
corrections to both lepton pair and Higgs boson production at the third order
(N$^3$LO) in the strong coupling constant $\as$, and extend the corresponding 
threshold resummations to the N$^3$LL contributions. The first step is achieved
by analysing the general mass-factorization structure of the third-order 
coefficient functions $\cd$ in terms of recent results on the three-loop 
splitting functions~\cite{Moch:2004pa,Vogt:2004mw} and the quark and gluon form
factors \cite{Moch:2005id,Moch:2005tm}, analogous to the second-order procedure
of Ref.~\cite{Matsuura:1987wt}. For the second step we just have to determine 
one resummation coefficient, usually denoted by $D_3$, from these results, as 
the structure of the resummation exponent has already been derived in Ref.\
\cite{Moch:2005ba} to the required accuracy. 
As discussed below our results on $\cd$ for Higgs production, suitably treated 
in Mellin $N$-space, provide a very good approximation to the complete N$^3$LO 
corrections, thus facilitating improved predictions for the cross sections at 
the {\sc Tevatron} and the LHC.

In the soft limit, i.e., retaining only contributions of the forms 
\beq
\label{eq:DDdef}
  \DD_{k} \; = \; \left[ \frac{\ln^{\,k} (1-x)}{1-x} \right]_+ \:\: ,
  \quad \delta(1-x) 
\eeq
to the coefficient functions, only the respective subprocesses $\, q\bar{q} 
\,\ra\, \gamma^{\,\ast} \,\ra\, l^+ l^-\,$ and $\,gg \,\ra\, H\,$ contribute to
the Drell-Yan process and Higgs boson production. In the latter case, the $Hgg$
vertex is an effective interaction in the limit of a heavy top quark,
\beq
\label{eq:LggH}
   {\cal L}_{\,\rm eff} \: = \: - \frac{1}{4} \: C_H \: H \,
   G^{\,a}_{\mu\nu} G^{\,a,\mu\nu} \:\: ,
\eeq
where $G^{\,a}_{\mu\nu}$ denotes the gluon field strength tensor, and the
prefactor $C_H$ includes all QCD corrections to the top-quark loop. This
coefficient is of order $\as$ and known up to N$^3$LO ($\as^{\,4}$) 
\cite{Chetyrkin:1997un}.
The analysis of the higher-order corrections in the heavy-top limit is 
justified by the agreement between this approximation and the full 
calculations at NLO \cite{Dawson:1990zj,Djouadi:1991tk,Spira:1995rr}.

The general structure of the expansion coefficients $W^{\rm b}_n$ of the bare
partonic cross section, 
\bea
\label{eq:Wbare}
  W^{\rm b} \; = \; \sum\limits_{n=0}^\infty\, \: \left(\ars^{\rm{\,b}} 
                    \right)^n \, W^{\rm b}_{n} 
  \:\: , \quad \ars \; \equiv \; \frac{\as}{4\pi} \:\: ,
% \\[-6mm] \nn
\eea
is given by
 
\bea
 W^{\rm b}_0
   &\:=\:& \delta(1-x) \nn \\[0.5mm]
 W^{\rm b}_1
   &\:=\:& 2\,\mbox{Re}\,{\cal F}_1\,\delta(1-x) + {\cal S}_1 \nn \\[0.5mm]
 W^{\rm b}_2
   &\:=\:& (2\,\mbox{Re}\,{\cal F}_2 + \left|{\cal F}_1\right|^2) \delta(1-x)
         + 2\,\mbox{Re}\, {\cal F}_1 {\cal S}_1 + {\cal S}_2 \nn \\[0.5mm]
 W^{\rm b}_3
   &\:=\:& (2\,\mbox{Re}\,{\cal F}_3 + 2 \left|{\cal F}_1 {\cal F}_2 \right|)\, \delta(1-x)
         + (2\,\mbox{Re}\,{\cal F}_2 + \left|{\cal F}_1\right|^2) {\cal S}_1 
         + 2\,\mbox{Re}\,{\cal F}_1 {\cal S}_2 + {\cal S}_3\: .
\label{eq:Wbexp}
\eea
Here ${\cal F}_n$ denotes the bare $n$-loop time-like quark or gluon form 
factor, calculated in dimensional regularization with $D = 4 - 2\ep$ and, 
as all quantities, expanded in terms of $\ars = \as/(4\pi)$.
The dependence of the pure real-emission contributions ${\cal S}_n$ on the
scaling variable $x = M^{\,2\!}/s$ is 
given by the $D$-dimensional +-distributions $f_{2n,\ep}$ defined by
\beq
\label{eq:Dplus}
  f_{k,\ep}(x) \; = \; \ep [\,(1-x)^{-1-k\ep}\,]_+
               \; = \; - {1 \over k}\, \delta(1-x) + \sum_{i=0}\,
                       {(-k \ep)^i \over i\, !}\,\ep\,\DD_{\,i} 
\eeq
with $\DD_{\,k}$ of Eq.~(\ref{eq:DDdef}). As appropriate for a parallel
tretament of the two processes, the coefficient function for Higgs production
is defined as in Ref.~\cite{Ravindran:2003um}, i.e., $C_H$ in Eq.~(\ref
{eq:LggH}) is kept as a prefactor.   

The expansion coefficients $W_k$ obtained from Eq.~(\ref{eq:Wbexp}) after 
renormalizing the coupling constant,
\beq
\label{eq:asren}
  \ars^{\,b} \; = \; \ars - {\beta_0 \over \epsilon} \ars^{\,2}\,
  + \left({\beta_0^2 \over \epsilon^2}
  - {1 \over 2} {\beta_1 \over \epsilon}\right) \ars^{\,3} + \dots \:\: ,
\eeq
and multiplying, for Higgs production, with the square of the renormalization 
constant~\cite{Kluberg-Stern:1975rs,Collins:1977yq}
\beq
\label{eq:GGren}
  Z_{G^{\,2}} \; = \; \left[ 1 - \beta(\as)/(\ars\ep) \right]^{-1} \:\: ,\quad
  \beta(\as) \; = \; - \beta_0 \,\ars^{\,2} - \beta_1 \,\ars^{\,3} 
  - \beta_2 \,\ars^{\,4} - \ldots \:\: ,
\eeq
of the operator $G^{\,a}_{\mu\nu} G^{\,a,\mu\nu}$ in Eq.~(\ref{eq:LggH}) obey
the Kinoshita-Lee-Nauenberg theorem~\cite{Kinoshita:1962xx,Lee:1964xx} and
the mass-factorization relations 
(putting $\mu^{\,2} = M_{\gamma^{\,\ast}\! ,\,H}^{\,2}$)
\bea
\label{eq:wt1}
  W_1 & = & \frac{2}{\ep}\,\gz
       \: + \: \co \: + \: \ep\, \ao \: + \: \ep^2 \bo \: ,
\\[0.5mm]
\label{eq:wt2}
  W_2 & = & \frac{1}{\ep^2}
  \left\{ \left( 2\,\gz - \beta_0 \right) \gz \right\}
  \: + \: \frac{1}{\ep} \left\{ \go + 2 \co \gz \right\}
% \nn \\ & & \mbox{}
  \: + \: \ct + 2 \ao \gz
  \: + \: \ep \left\{ \at + 2 \bo \gz \right\} \: ,
\\[0.5mm]
\label{eq:wt3}
  W_3 & = & \frac{1}{3\ep^3}
  \left\{ 2 \left( \gz - \beta_0 \right)
  \left( 2\,\gz - \beta_0 \right) \gz \right\}
  \: + \: \frac{1}{3\ep^2} \left\{ 6\, \go \gz - 2 \beta_0 \go
  - 2 \beta_1 \gz  \right.
\nn \\[-0.5mm] & & \left. \mbox{}
   + 3\co \left( 2\,\gz - \beta_0 \right) \gz \right\}
  \: + \: \frac{1}{3\ep}\,\left\{ 2\, \gt + 3 \co\,\go + 6 \ct\,
     \gz + 3 \ao \left( 2\,\gz - \beta_0 \right) \gz \right\}
\nn \\[1.0mm] & & \mbox{}
  \: + \: \cd + \ao \go + 2 \at \gz +
     \bo \left( 2\,\gz - \beta_0 \right) \gz \:\: .
\eea
Here the anomalous dimensions $\gamma_k^{}$ are related by a conventional
sign to the diagonal Altarelli-Parisi splitting functions $P_k$. 
In $x$-space, where these quantities (in the \MSb\ scheme adopted throughout 
this article) have soft limits of the form \cite{Korchemsky:1989si}
\beq
\label{eq:Psoft}
  P_{k-1}^{} \; = \; A_k\, \DD_0 + P_k^{\,\delta}\: \delta (1-x) \:\: ,
\eeq
the products in Eq.~(\ref{eq:wt1}) -- (\ref{eq:wt3}) have to be read as Mellin 
convolutions. To the required accuracy these convolutions can be readily 
carried out using, for example, the appendix of Ref.~\cite{vanNeerven:2001pe}.

Eqs.~(\ref{eq:wt1}) -- (\ref{eq:wt3}) can be used to derive all $\DD_k$
contributions to the \MSb\ coefficient function $c_{n\,}^{}$, once the
coefficients $A_n$, $P_n^{\,\delta}$ are known together with all $1/\ep$ pole
terms of the $n$-loop form factor and suitable lower-order information.  The
salient point for this extraction is the structure (\ref{eq:Dplus}) of the soft
emissions linking the coefficients of $\ep^{-1}\,\delta(1-x)$ to those of
$\DD_0$; thus a mass-factorization constraint on the former term fixes the
latter coefficient.
With the results of Refs.~\cite
{Matsuura:1989sm,Ravindran:2003um,Ravindran:2004mb} and
\cite{Moch:2004pa,Vogt:2004mw,Moch:2005id,Moch:2005tm} the above conditions
are fulfilled at $n=3$ for both lepton pair and Higgs boson production.

Thus, treating Higgs boson production first, we insert the $\ep$-expansion 
\beq
  {\cal S}_n \; = \; f_{2n,\ep}\: \sum_{l=-2n}^{\infty}\: 
  2n\: s_{n,l}^{}\: \ep^{\,l} 
\eeq
into Eqs.~(\ref{eq:wt1}) -- (\ref{eq:wt3}) and recursively determine the
coefficients $s_{n,l}^{}$. The first-order result is known to all orders in 
$\ep$. For later convenience, we here present its expansion up to order 
$\ep^3$,
\beq
\label{eq:Sexp}
  {\cal S}_1 \; = \; 2\, f_{2,\ep}\: C_A \left\{ \mbox{}
  - \frac{4}{\ep^2} + 6\,\z2 + \frac{28}{3}\,\z3\: \ep 
  + \frac{3}{2}\,\zs\: \ep^2 + \ep^3 \left[ - 14\,\z2\z3 
     + \frac{124}{5}\,\z5 \right] + \ldots
  \right\} \:\: .
\eeq
The corresponding second-order coefficients read
\bea
  s_{2,-4}^{} &\! =\! & - 8\, \cas
\nn \\[1mm]
  s_{2,-3}^{} &\! =\! & - {11 \over 3}\, \cas + {2 \over 3}\, \ca\nf
\nn \\[1mm]
  s_{2,-2}^{} &\! =\! & 
      \cas \left[ - {67 \over 9} + 58\,\z2 \right]
    + {10 \over 9}\, \ca \nf
\nn \\[1mm]
  s_{2,-1}^{} &\! =\! & 
      \cas \left[  - {404 \over 27} + {77 \over 3}\,\z2 
                   + {538 \over 3}\,\z3 \right] 
    + \ca\nf \left[{56 \over 27} - {14 \over 3}\,\z2 \right]
\nn \\[1mm]
  s_{2,0}^{}\;\;&\! =\! & 
      \cas \left[ - {2428 \over 81} + {469 \over 9}\,\z2 
    + {682 \over 9}\,\z3 + {16 \over 5}\,\zs \right]
    + \ca\nf \left[ {328 \over 81} - {70 \over 9}\,\z2 
                  - {124 \over 9}\,\z3 \right] \: .
\label{eq:Hsoft2coeffs}
\eea
Here $\nf$ denotes the number of effectively massless quark flavours, $C_F$
and $C_A$ are the usual SU(N) colour factors, with $C_F= 4/3$ and $C_A= 3$ 
for QCD, and $\zeta_n$ represents Riemann's $\zeta$-function.

From $s_{2,-4}^{}\,\ldots\, s_{2,-1}^{}$ and the first-order results one 
recovers the $\DD_k$ terms \cite{Harlander:2001is,Catani:2001ic} of the NNLO 
coefficient function at the renormalization and factorization scales 
$\mu_r^{\,2} = \mu_{\! f}^{\,2} = M_{H}^{\,2}$,
\bea
\label{eq:c2H}
  \ct(x) &\! =\! & 128\,\cas\,\DD_3 
  - \left\{ 
         {176 \over 3}\, \cas
       - {32 \over 3}\, \ca\nf 
    \right\} \DD_2
  + \left\{
         \cas \left[ {1072 \over 9} - 160\,\z2 \right] 
       - {160 \over 9} \ca\nf
    \right\} \DD_1
\nn \\[1mm] & & \mbox{}
  + \left\{
         \cas \left[ - {1616 \over 27} + {176 \over 3}\,\z2 + 312\,\z3 \right]
       + \ca\nf \left[ {224 \over 27} - {32 \over 3}\,\z2\right] 
    \right\} \DD_0
\nn \\[1mm] & & \mbox{}
  + \left\{
         \cas \left[ 93 + {536 \over 9}\,\z2 - {220 \over 3}\,\z3 
         - {4 \over 5}\,\zs \right]           
       - \ca\nf \left[ {80 \over 3} + {80 \over 9}\,\z2 + {8 \over 3}\,\z3 
         \right] 
\right.
\nn \\[1mm] & & \left. \quad \mbox{} 
       - \cf\nf \left[ {67 \over 3} - 16\,\z3 \right] 
    \right\} \delta(1-x) 
\:\: .
\eea
The SU(N) result (\ref{eq:Hsoft2coeffs}) for the  coefficient $s_{2,0}^{}$, on 
the other hand, is fixed by the corresponding $\delta(1-x)$ contribution to 
Eq.~(\ref{eq:c2H}) derived in Ref.~\cite{Ravindran:2004mb}. Actually this
colour-factor decomposition is checked (and could have been predicted from
the ${\rm N} = 3$ QCD results of Refs.~\cite{Harlander:2001is,Catani:2001ic})
by the absence of a $C_F \nf$ term in ${\cal S}_2$ obvious from the colour 
structure of the contributing Feynman diagrams. 

The quantity $\at$ in Eq.~(\ref{eq:wt2}) has not been computed so far, hence
the coefficient $s_{2,1}^{}$ is unknown at this point. This suggests a problem,
as this coefficient enters the $\ep^0$ part of Eq.~(\ref{eq:wt3}).
However, its contribution to the $\DD_0$ term (but not the $\delta(1-x)$ piece)
of $\cd$ is found to cancel in the end. Keeping $s_{2,1}^{}$ as an unknown
in the intermediate relations, the third-order coefficients in 
Eq.~(\ref{eq:Sexp})~are
\bea
  s_{3,-6}^{} &\! =\! & - {32 \over 3}\, \cat
\nn \\[1mm]
  s_{3,-5}^{} &\! =\! & - {44 \over 3}\,\cat + {8 \over 3}\,\cas\nf
\nn \\[1mm]
  s_{3,-4}^{} &\! =\! & 
      \cat \left[ - {2896 \over 81} + 184\,\z2 \right] 
    + {536 \over 81}\cas\nf - {16 \over 81}\,\ca\nsq
\nn \\[1mm]
  s_{3,-3}^{} &\! =\! & 
      \cat \left[ - {21052 \over 243} + {6710 \over 27}\,\z2
                  + {2440 \over 3}\,\z3 \right]
    + \cas\nf \left[ {4148 \over 243} - {1220 \over 27}\,\z2 \right] 
    + {4 \over 9}\, \ca\cf\nf - {160 \over 243}\, \ca\nsq \!\!\!
\nn \\[1mm]
  s_{3,-2}^{} &\! =\! &
      \cat \left[ - {51322 \over 243} + {48856 \over 81}\,\z2
                  + {29876 \over 27}\,\z3 - {7592 \over 45}\,\zs \right]
    + \ca\cf\nf \left[ {110 \over 27} - {32 \over 9}\,\z3 \right]
\nn \\[0.5mm] & & \mbox{\hspace*{-4mm}}
    + \cas\nf \left[ {10468 \over 243} - {9004 \over 81}\,\z2
                   - {5336 \over 27}\,\z3 \right]
    + \ca\nsq \left[ - {16 \over 9} + {88 \over 27}\,\z2 \right]
\nn \\[1mm]
  s_{3,-1}^{} &\! =\! &
      \cat \left[ - {617525 \over 2187} + {251942 \over 243}\,\z2
                  + {56032 \over 27}\,\z3 - {1661 \over 10}\,\zs
                  - {79388 \over 9}\,\z2\z3 + {49888 \over 5}\,\z5 \right]
\nn \\[0.5mm] & & \mbox{\hspace*{-4mm}}
    + \ca\cf\nf \left[ {1711 \over 81} - {22 \over 3}\,\z2
                     - {304 \over 27}\,\z3 - {32 \over 15}\,\zs \right]
    + \ca\nsq \left[ - {9728 \over 2187} + {880 \over 81}\,\z2
                   + {1040 \over 81}\,\z3 \right]
\nn \\[0.5mm] & & \mbox{\hspace*{-4mm}}
    + \cas\nf \left[ {164194 \over 2187} - {55154 \over 243}\,\z2
                   - {31520 \over 81}\,\z3 + {97 \over 3}\,\zs \right]
    + 4\,\ca s_{2,1}^{} \:\: .
\label{eq:Hsoft3coeffs}
\eea
Analogous to $s_{2,0}^{}$ discussed above,
the coefficient $s_{3,0}^{}$ cannot be derived by mass-factorization
arguments, but requires a third-order calculation 
like in the case of deep-inelastic scattering~\cite{Vermaseren:2005qc}.

The above results, after combination with the gluon splitting function 
\cite{Vogt:2004mw} and form factor~\cite{Moch:2005tm} according to
Eqs.~(\ref{eq:Wbare}) and (\ref{eq:Wbexp}), lead to the following soft-emission
contribution to the third-order (N$^3$LO) coefficient function for Higgs boson 
production at $\mu_r^{\,2} = \mu_{\! f}^{\,2} = M_H^{\,2}$:
\bea
\label{eq:c3H5}
  \cd \Big|_{\,\DD_5} &\! =\! & 512\,\cat 
\:\: ,\\[1mm] 
\label{eq:c3H4}
  \cd \Big|_{\,\DD_4} &\! = & - {7040 \over 9}\,\cat 
    + {1280 \over 9}\,\cas\nf
\:\: ,\\[2mm] 
\label{eq:c3H3}
  \cd \Big|_{\,\DD_3} &\! =\! &
      \cat \left[ {59200 \over 27} - 3584\,\z2 \right]
    - {10496 \over 27}\,\cas\nf + {256 \over 27}\, \ca\nsq
\:\: ,\\[2mm]
\label{eq:c3H2}
  \cd \Big|_{\,\DD_2} &\! =\! &
      \cat \left[ - {67264 \over 27} + {11968 \over 3}\,\z2
                  + 11584\,\z3 \right]
    + \cas\nf \left[ {14624 \over 27} - {2176\over 3}\,\z2 \right]
\nn \\[0.5mm] & & \mbox{\hspace*{-4mm}}
    + 32\,\ca\cf\nf - {640 \over 27}\, \ca\nsq
\:\: ,\\[2mm]
\label{eq:c3H1}
  \cd \Big|_{\,\DD_1} &\! =\! &
    \cat \left[ {244552 \over 81} - {9728 \over 3}\,\z2
                  - {22528 \over 3}\,\z3 - {9856 \over 5}\,\zs \right]
    - \ca\cf\nf \left[ 504 - 384\,\z3 \right]
\nn \\[0.5mm] & & \mbox{\hspace*{-4mm}}
    + \cas\nf \left[ - {67376 \over 81} + {6016 \over 9}\,\z2
                     + {2944 \over 3}\,\z3 \right]
    + \ca\nsq \left[ {1600 \over 81} - {256 \over 9}\,\z2 \right]
\:\: ,\\[2mm]
\label{eq:c3H0}
  \cd \Big|_{\,\DD_0} &\! =\! &
      \cat \left[ - {594058 \over 729} + {137008 \over 81}\,\z2
                  + {143056 \over 27}\,\z3 + {4048 \over 15}\,\zs
                  - {23200 \over 3}\,\z2\z3 + 11904\,\z5 \right]
\nn \\[0.5mm] & & \mbox{\hspace*{-4mm}}
    + \cas\nf \left[ {125252 \over 729} - {34768 \over 81} \,\z2
                 - {7600 \over 9}\,\z3 - {544 \over 15}\,\zs \right]
\\[0.5mm] & & \mbox{\hspace*{-4mm}}
    + \ca\cf\nf \left[ {3422 \over 27}  -32\,\z2
                 - {608 \over 9}\,\z3 - {64 \over 5}\,\zs \right]
    - \ca\nsq \left[ {3712 \over 729} - {640 \over 27}\,\z2
                 - {320 \over 27}\,\z3 \right]
\nn \:\: .
\eea
Eqs.~(\ref{eq:c3H5}) -- (\ref{eq:c3H1}) agree with the results derived from 
the NNLL threshold resummation in Ref.~\cite{Catani:2003zt}. 
Eq.~(\ref{eq:c3H0}) represents a new result of the present study. 

We now turn to lepton pair production. The corresponding coefficients 
$s_{n,l}^{}$ can be derived in the same manner, some being related to 
their Higgs counterparts by simple replacements of $C_F$ by $C_A$.
Specifically, ${\cal S}_1$ for the Drell-Yan case is obtained from 
Eq.~(\ref{eq:Sexp}) by a factor $C_F/C_A$, and $s_{n,-2n+k}^{}$ for 
$n \geq 2$ and $k = 0,\: 1$ by multiplication with $(C_F/C_A)^{n-k}$
from Eqs.~(\ref{eq:Hsoft2coeffs}) and (\ref{eq:Hsoft3coeffs}).
Accordingly the coefficients of $\DD_5$ and $\DD_4$ of the third-order 
coefficient function for the Drell-Yan process are related to 
Eqs.~(\ref{eq:c3H5}) and (\ref{eq:c3H4}), respectively, by factors 
$\cft/\cat$ and $\cfs/\cas$. The remaining coefficients for the 
standard scale choice $\mu_r^{\,2} = \mu_{\! f}^{\,2} = 
M_{\,\gamma^{\,\ast}}^{\,2}$ are given by
\bea
\label{eq:c3dy3}
  \cd \Big|_{\,\DD_3} &\! =\! & 
      {7744 \over 27}\,\cas\cf 
    + \ca\cfs \left[ {17152 \over 9} - 512\,\z2 \right]
    - \cft \left[ 2048 + 3072\,\z2 \right]
    - {2816 \over 27}\, \ca\cf\nf 
\nn \\[0.5mm] & & \mbox{\hspace*{-4mm}}
    - {2560 \over 9}\, \cfs\nf
    + {256 \over 27} \cf\nsq
\:\: ,\\[2mm] 
\label{eq:c3dy2}
  \cd \Big|_{\,\DD_2} &\! =\! & 
      \cas\cf \left[ - {28480 \over 27} + {704 \over 3}\,\z2 \right]
    - \ca\cfs \left[ {4480 \over 9} - {11264 \over 3}\,\z2
                   - 1344\,\z3 \right]
    + 10240\,\z3\: \cft
\nn \\[0.5mm] & & \mbox{\hspace*{-4mm}}
    + \ca\cf\nf \left[ {9248 \over 27} - {128 \over 3}\,\z2 \right]
    + \cfs\nf \left[ {544 \over 9} - {2048 \over 3}\,\z2 \right]
    - {640 \over 27} \cf\nsq 
\:\: ,\\[2mm] 
\label{eq:c3dy1}
  \cd \Big|_{\,\DD_1} &\! =\! &
      \cas\cf \left[ {124024 \over 81} - {12032 \over 9}\,\z2 
                 - 704\,\z3 + {704 \over 5}\,\zs \right]
\nn \\[0.5mm] & & \mbox{\hspace*{-4mm}}
    - \ca\cfs \left[ {35572 \over 9} + {11648 \over 9}\,\z2
                 + 5184\,\z3 - {3648 \over 5}\,\zs \right]
\nn \\[0.5mm] & & \mbox{\hspace*{-4mm}}
    + \cft \left[ 2044 + 2976\,\z2 - 960\,\z3 - {14208 \over 5}\,\zs \right]
    - \ca\cf\nf \left[ {32816 \over 81} - 384\,\z2 \right]
\nn \\[0.5mm] & & \mbox{\hspace*{-4mm}}
    + \cfs\nf \left[ {4288 \over 9} + {2048 \over 9}\,\z2 
                 + 1280\,\z3 \right]
    + \cf\nsq \left[ {1600 \over 81} - {256 \over 9}\,\z2 \right]
\:\: ,\\[2mm] 
\label{eq:c3dy0}
  \cd \Big|_{\,\DD_0} &\! =\! &
      \cas\cf \left[ - {594058 \over 729} + {98224 \over 81}\,\z2
                 + {40144 \over 27}\,\z3 - {2992 \over 15}\,\zs 
                 - {352 \over 3}\,\z2\z3 - 384\,\z5 \right]
\nn \\[0.5mm] & & \mbox{\hspace*{-4mm}}
    + \ca\cfs \left[ {25856 \over 27} - {12416 \over 27}\,\z2
                 + {26240 \over 9}\,\z3 + {1408 \over 3}\,\zs 
                 - 1472\,\z2\z3 \right]
\nn \\[0.5mm] & & \mbox{\hspace*{-4mm}}
    - \cft \Big[ 4096\,\z3 + 6144\,\z2\z3 - 12288\,\z5 \Big]
    - \cf\nsq \left[ {3712 \over 729} - {640 \over 27}\,\z2 
                 - {320 \over 27}\,\z3 \right]
\nn \\[0.5mm] & & \mbox{\hspace*{-4mm}}
    + \ca\cf\nf \left[ {125252 \over 729} - {29392 \over 81} \,\z2 
                 - {2480 \over 9}\,\z3 + {736 \over 15}\,\zs \right]
\nn \\[0.5mm] & & \mbox{\hspace*{-4mm}}
    - \cfs\nf \left[ 6  - {1952 \over 27}\,\z2
                 + {5728 \over 9}\,\z3 + {1472 \over 15}\,\zs \right]
\:\: .
\eea
Also Eqs.~(\ref{eq:c3dy3}) -- (\ref{eq:c3dy1}) agree with the results derived 
from the NNLL threshold resummation~\cite{Vogt:2000ci}, while 
Eq.~(\ref{eq:c3dy0}) is a new result of this study.
The additional coefficients for $\mu_{\! f}^{\,2} \neq M_{\,\gamma^{\,\ast}}
^{\,2}$ and $\mu_r^{\,2} \neq \mu_{\! f}^{\,2}$ can be derived analogously to
Eqs.~(2.16) -- (2.18) of Ref.~\cite{vanNeerven:2000uj} or using the N$^3$LL
threshold resummation expression~\cite{Moch:2005ba}, but will be skipped here
for brevity.

We note that the $\zeta$-function terms of highest transcendentality $n$, i.e, 
the coefficients of $\zeta_n$ and $\zeta_i\,\zeta_j$ with $\,i+j=n$, in the 
$\ep^{-2l+n}$ contributions to the pure real-emission function ${\cal S}_l$ 
agree between Higgs production and the Drell-Yan process for the 
Super-Yang-Mills case $C_A = C_F = n_c$. The same holds for the quark and gluon
form factors \cite{Moch:2005id,Moch:2005tm} and, consequently, also for the 
$\zeta$-function terms of weight $n$ in the soft logarithms $\DD_{2l-1-n}^{}$ 
of the coefficient functions for Higgs boson and lepton pair production, see
Eqs.~(\ref{eq:c3H5})--(\ref{eq:c3H0}) and (\ref{eq:c3dy3})--(\ref{eq:c3dy0}).
By construction, generalizing Eq.~(\ref{eq:Wbexp}), this feature extends to all
orders of perturbation theory.

We now turn to the threshold resummation. For the processes under 
consideration, the coefficient functions exponentiate after transformation 
to Mellin $N$-space~\cite{Sterman:1987aj,Catani:1989ne},
\beq
\label{eq:cNres}
  C^{\,N} \; =\; ( 1 + \ars\,g_{01}^{} + \ars^{\,2}\,g_{02}^{} + \ldots )
  \cdot \exp\, (G^N) \: + \: {\cal O}(N^{-1}\ln^n N) \:\: .
\eeq
Here $g_{0k}^{}$ collects the $N$-independent contributions at the $k$-th 
order, and the resummation exponent $G^N$ takes the form
\beq
\label{eq:GNexp}
  G^N(Q^2) \: = \:
  \ln N \cdot g_1^{}(\lambda) \: + \: g_2^{}(\lambda) \: + \:
  \ars\, g_3^{}(\lambda) \: + \: \ars^{\,2}\, g_4^{}(\lambda) \: + \:
  \ldots 
\eeq
with $\lambda = \beta_0\, \ars\, \ln\, N$. The functions $g_3^{}$ and $g_4^{}$
have been determined in Refs.~\cite{Vogt:2000ci,Catani:2003zt} and 
\cite{Moch:2005ba}, respectively. Besides the quantities $A_k$ in Eq.~(\ref{eq:Psoft}) 
and lower-order coefficients, the functions $g_k$ depend on one parameter, 
usually denoted by $D_{k-1}$.

Before we present our new results for the coefficient $D_3$, we recall, for 
the convenience of the reader, the $N$-independent first- and second-order 
contributions which enter its determination. 
For Higgs boson production these coefficients read 
\bea
\label{eq:g01H}
  g_{01}^{} &\! =\! & \ca ( 16\,\z2 + 8\,\ecs )
\:\: ,\\[2mm]
\label{eq:g02H}
  g_{02}^{} &\! =\! &
    \cas \left[ 93 + {1072 \over 9}\,\z2 - {308 \over 9}\,\z3
              + 92\,\zs + {1616 \over 27}\,\ec - 56\,\ec\z3
              + {536 \over 9}\,\ecs + 112\,\ecs \z2 \right.
\nn \\[0.5mm] & & \left. \mbox{}
              + {176 \over 9}\,\ect + 32\,\ecf \right]
  + \ca\nf \left[ - {80 \over 3} - {160 \over 9}\,\z2 - {88 \over 9}\,\z3
                  - {224 \over 27}\,\ec - {80 \over 9}\,\ecs 
                  - {32 \over 9}\,\ect \right]
\nn \\[0.5mm] & & \mbox{\hspace*{-4mm}}
  + \cf\nf \left[ - {67 \over 3} + 16\,\z3 \right] \:\: .
\eea
The corresponding results for the Drell-Yan case are given by 
\bea
\label{eq:g01DY}
  g_{01}^{} &\! =\! & \cf ( -16 + 16\,\z2 + 8\,\ecs )
\:\: ,\\[1mm]
\label{eq:g02DY}
  g_{02}^{} &\! =\! & 
    \cfs \left[ {511 \over 4} - 198\,\z2 - 60\,\z3 + {552 \over 5}\,\zs 
              - 128\,\ecs + 128\,\ecs \z2 + 32\,\ecf \right]
\nn \\[0.5mm] & & \mbox{\hspace*{-4mm}}
  + \ca\cf \left[ - {1535 \over 12} + {376 \over 3}\,\z2 + {604 \over 9}\,\z3 
                  - {92 \over 5}\,\zs + {1616 \over 27}\,\ec
                  - 56\,\ec \z3 + {536 \over 9}\,\ecs \right.
\nn \\[0.5mm] & & \left. \mbox{}
                  - 16\,\ecs \z2 + {176 \over 9}\,\ect \right]
  + \cf\nf \left[ {127 \over 6} - {64 \over 3}\,\z2 + {8 \over 9}\,\z3 
                - {224 \over 27}\,\ec - {80 \over 9}\,\ecs 
                - {32 \over 9}\,\ect \right] \:\: . \quad
\eea

Inserting the above results into the explicit formulae for the resummation
exponents~\cite{Vogt:2000ci,Catani:2003zt,Moch:2005ba}, we recover the known
coefficients 
(as always referring to the expansion parameter $\ars = \as/(4\pi)^{\,}$)
\bea
\label{eq:D1}
  D_1 &\! = \! & 0
\:\: ,\\[1mm]
\label{eq:D2}
  D_2 &\! = \! & 
   C_I \left[
   C_A \left( - \frac{1616}{27} + \frac{176}{3}\,\z2 + 56\,\z3 \right)
   \: + \: \nf \left( \frac{224}{27} - \frac{32}{3}\,\z2 \right)
   \right] \:\: ,
\eea
and derive the new third-order contribution
\bea
\label{eq:D3}
  D_3 & = & 
      C_I\,\cas \left[ - {594058 \over 729} + {98224 \over 81}\,\z2
                   + {40144 \over 27}\,\z3 - {2992 \over 15}\,\zs
                   - {352 \over 3}\,\z2\z3 - 384\,\z5 \right]         
\nn \\[0.5mm] & & \mbox{\hspace*{-4mm}}
    + C_I\,C_A\nf  \left[ {125252 \over 729} - {29392 \over 81}\,\z2
                   - {2480 \over 9}\,\z3 + {736 \over 15}\,\zs \right]
\\[0.5mm] & & \mbox{\hspace*{-4mm}}
    + C_I\,C_F\nf \left[ {3422 \over 27} - 32\,\z2
                   - {608 \over 9}\,\z3 - {64 \over 5}\,\zs \right]
    + C_I\,\nsq \left[ - {3712 \over 729} + {640 \over 27} \,\z2
                   + {320 \over 27}\,\z3 \right]
\nn
\eea
with $C_I = C_F$ for the Drell-Yan case, and $C_I = C_A$ for Higgs production.
Hence, not unexpectedly, we find that also $D_3$ is maximally non-abelian, 
with the quark and gluon cases related by an overall factor $C_F/C_A$. This
is the same behaviour as shown by the cusp anomalous dimensions $A_n$ in 
Eq.~(\ref{eq:Psoft})~\cite{Korchemsky:1989si} and by the form-factor
resummation coefficients $f_n$ known up to three loops~\cite{Ravindran:2004mb,%
Moch:2005tm}. In fact, there is a simple relation between the coefficients 
$D_n$ and $f_n$ (using the notation of Ref.~\cite{Moch:2005tm}),
\bea
  D_2 & = & 2\beta_0 \,s_{1,0}^{} - 2 f_2
\nn \\[1mm] 
  D_3 & = & 2\beta_1 \,s_{1,0}^{} - 4\beta_0^2\, s_{1,1}^{}
  + 4 \beta_0 \left( s_{2,0}^{} - 36/5\,\zs\,C_I^{\,2} \right) - 2 f_3 
  \:\: ,
\label{eq:Dvsf}
\eea
of which the first line of has been derived before in Ref.~\cite{Eynck:2003fn} 
from an extension of the threshold resummation to $N$-independent 
contributions. In the present mass-factorization framework, the $s_{n,l}^{}$ 
terms in Eqs.~(\ref{eq:Dvsf}) can be traced back to the coupling-constant 
renormalization of Eqs.~(\ref{eq:Wbexp}).

We now turn to the numerical impact of the N$^3$LO and resummation corrections
to the coefficient functions, confining ourselves to Higgs boson production for
brevity. All parameters are taken over from Ref.~\cite{Ravindran:2003um}, i.e.,
we use the values $m_{\,t}=173.4$ GeV (very close to the present world average)
and $G_F= 4541.7$ pb for the top mass and
Fermi constant in the prefactor $C_H$ in Eq.~(\ref{eq:LggH}), and the parton 
distributions of Refs.~\cite{Martin:2001es,Martin:2002dr} with their respective
values of strong coupling constant at LO, NLO and NNLO, $\as(M_Z)$ = 0.130, 
0.119 and 0.115. Anticipating a slight further reduction at N$^3$LO, we employ
$\as(M_Z)$ = 0.114 at this order. The N$^3$LO corrections to $C_H$ in the
heavy-top limit are taken from Ref.~\cite{Chetyrkin:1997un}. All higher-order
contributions are calculated in the heavy top-quark approximation, but 
normalized to the full lowest-order result.

As mentioned above, the $\delta(1-x)$ contributions to the N$^3$LO coefficient
functions $\cd$ cannot be derived at this point. However, we note that the 
coefficients $g_{0k}^{}$ in Eqs.~(\ref{eq:g01H}) -- (\ref{eq:g02DY}) are much
larger than their $\delta(1-x)$ counterparts for $\co$ and especially for 
$\ct$. We expect the same behaviour for $\cd$. Moreover, a good approximation 
(to about 10\% or less) to the full double convolutions $g\ast g\ast [c_i(x)/x]$
is obtained at NLO and NNLO by transforming to $N$-space and keeping only the
$\ln^k N$ and $N^0$ terms arising from the +-distributions (but not the
$\delta$-function) in $\co$ and $\ct$.
Consequently Eqs.~(\ref{eq:c3H5}) -- (\ref{eq:c3H0}) facilitate a sufficient
approximation to the complete~N$^3$LO~correction, to which we assign a 
conservative 20\% uncertainty, i.e., twice the offset found at NLO and NNLO. 

\begin{figure}[bhtp]
\centerline{\epsfig{file=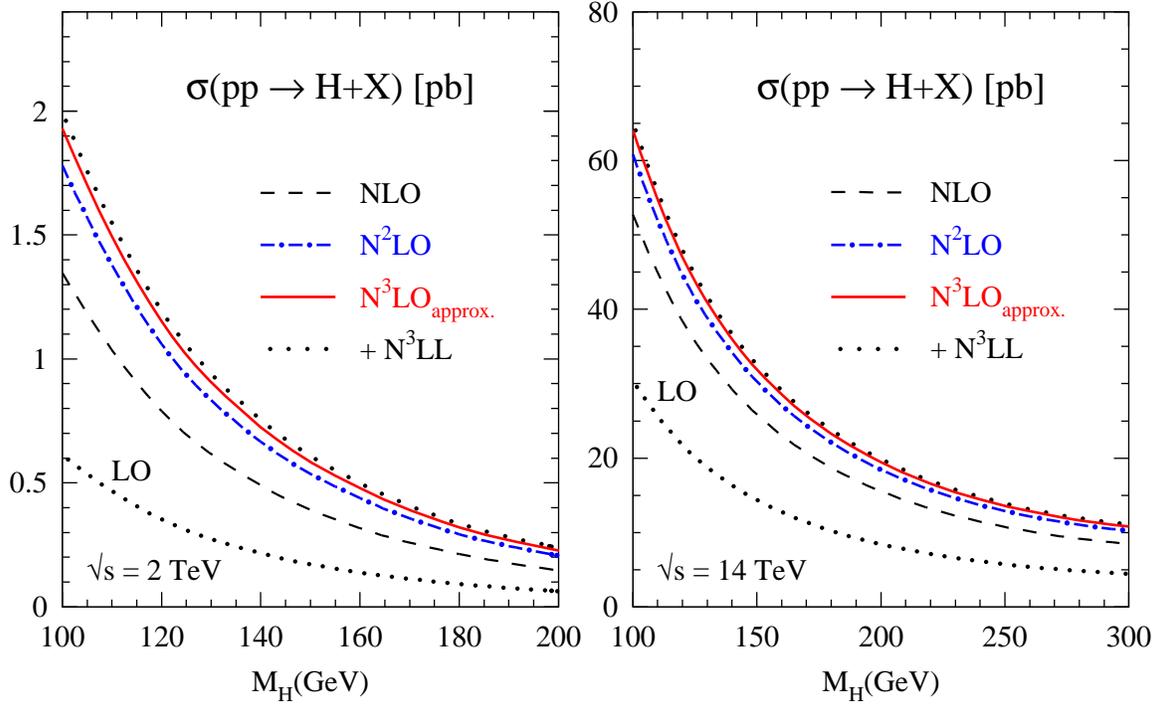,width=15.5cm,angle=0}}
\vspace{-2mm}
\caption{ \label{pic:fig1}
 The perturbative expansion of the total cross section for Higgs boson
 production at the {\sc Tevatron} (left) and the LHC (right) for the standard 
 scale choice $\,\mu_r = \mu_{\! f}^{} = M_H$. }
\vspace*{-2mm}
\end{figure}

\begin{figure}[thbp]
\centerline{\epsfig{file=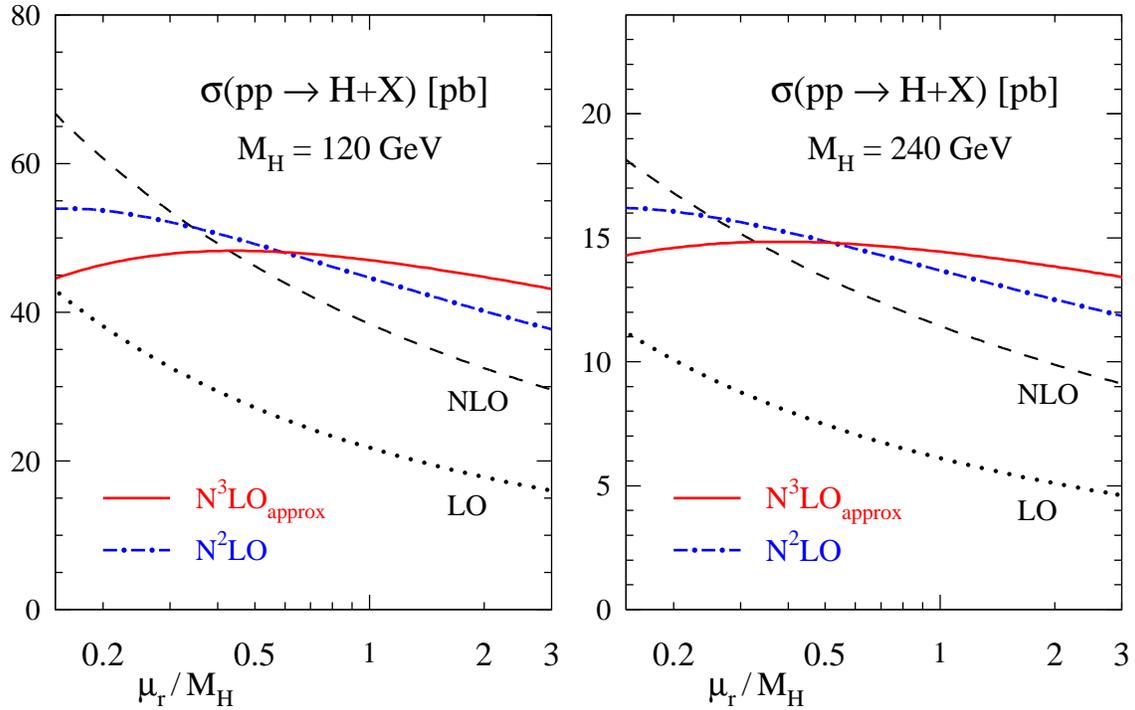,width=15.5cm,angle=0}}
\vspace{-2mm}
\caption{ \label{pic:fig2}
 The dependence of the fixed-order predictions for the LHC cross section
 on the renormalization scale $\mu_r$ at $\,\mu_{\! f}^{} = M_H\,$ for 
 two representative values of the Higgs boson mass $M_H$.}
\vspace{-3.5mm}
\end{figure}

The corresponding results are added in Fig.~\ref{pic:fig1} to the total cross
sections up to NNLO \cite{Harlander:2002wh,Anastasiou:2002yz,Ravindran:2003um}
at the {\sc Tevatron} and the LHC for the standard choice  $\,\mu_r =
\mu_{\! f}^{} = M_H\,$ of the renormalization and factorization scales.
The dependence of the cross sections on $\mu_r$ is illustrated in 
Fig.~\ref{pic:fig2} for two representative values of the Higgs mass $M_H$.
Also shown in Fig.~\ref{pic:fig1} are the additional contributions of the 
N$^3$LL threshold resummation (\ref{eq:GNexp}), see also Ref.~\cite
{Moch:2005ba}, of the terms beyond N$^3$LO.
In principle this resummation requires a second coefficient, the four-loop cusp
anomalous dimension $A_4$, besides $D_3$ of Eq.~(\ref{eq:D3}). However, the 
effect of $A_4$ can safely be expected to be very small, as corroborated by the
Pad\'e estimate of Ref.~\cite{Moch:2005id} employed in our numerical analysis. 
The Mellin-inversion of the exponentiated result (which is entirely dominated 
by the next few orders in $\as$ in the present case) has been performed using 
the standard `minimal prescription' contour~\cite{Catani:1996yz}.  
 
The inclusion of our new result for the coefficient function $\cd$ effects
an increase of the cross sections by about 10\% at the {\sc Tevatron} and 5\%
at the LHC. The estimated uncertainties due to the approximate character of 
$\cd$ (see above) thus amount to 2\% and 1\%, respectively. 
Contributions of yet higher orders, as estimated by the threshold resummation, 
lead to a further increase by roughly half the N$^3$LO effect.
Lacking N$^3$LO and threshold-resummed (see Ref.~\cite{Corcella:2005us} for a 
first study) parton distributions, the NNLO gluon distribution of 
Ref.~\cite{Martin:2002dr} has been employed for all results beyond the 
next-to-leading order. Based on the pattern of the available orders, one may
expect slightly smaller (by about 2\%) gluon-gluon luminosities at N$^3$LO.
    
The  residual uncertainty due to uncalculated contributions of yet higher order
is often estimated by varying $\mu_r$ and$\,$/$\,$or $\mu_{\!f}^{}$.
At the LHC the representative variation of $\mu_r$ with fixed $\mu_{\!f}^{}$, 
illustrated in Fig.~\ref{pic:fig2} for two Higgs masses, yields uncertainties 
of less than 4\% for the conventional interval $1/2\: M_H \leq \mu_r \leq 2\,
M_H$ at N$^3$LO, an improvement by almost a factor of three over the 9 to 11\% 
stability of the NNLO cross sections.
At the {\sc Tevatron} the corresponding $\mu_r$ dependence decreases from about
11\% at NNLO to 5\% at N$^3$LO for $\,M_H = 120$ GeV where, as in Fig.~\ref
{pic:fig2}, the N$^3$LO cross section exhibits a stationary point close to 
$\,\mu_r = 1/2\: M_H$.
Considering these and the above results, 5\% at the LHC and 7\% at the 
{\sc Tevatron} appear to represent conservative estimates of the improved 
cross-section uncertainties due to the truncation of the perturbation series 
at N$^3$LO.

\vspace*{2mm}
\noindent
{\bf Acknowledgments:} 
We are grateful to E. Laenen for a brief, but very stimulating discussion,
and to J. Smith, J. Bl\"umlein and V. Ravindran for providing the {\sc Fortran}
codes of Refs.~\cite{Ravindran:2003um} and \cite{Blumlein:2005im}.
The symbolic manipulations for this study have been performed in {\sc Form}
\cite{Vermaseren:2000nd}.
The work of S.M. has been supported in part by the Helmholtz Gemeinschaft
under contract VH-NG-105 and by the Deutsche Forschungsgemeinschaft in
Sonderforschungs\-be\-reich/Transregio~9.

\vspace*{2mm}
\noindent
{\bf Note added:}
Shortly after the completion of this letter, Ref.~\cite{Laenen:2005uz} appeared,
which addresses the threshold resummation especially for lepton-pair production
in the approach of Ref.~\cite{Eynck:2003fn}. In particular, our result 
(\ref{eq:D3}) for the coefficient $D_3$ for the Drell-Yan process is confirmed 
by this research.

\end{document}